\documentclass[useAMS,usenatbib]{mn2e}
\usepackage{graphicx}
\usepackage{amssymb}

\title[Convection in Hyperaccretion Disks]{Possible Origin of Rapid Variability of Gamma-Ray Bursts due to Convective Energy Transfer in Hyperaccretion Disks}

\author[N. Kawanaka \& K. Kohri]{Norita Kawanaka$^{1}$\thanks{E-mail:norita.kawanaka@kek.jp} \& Kazunori Kohri$^{1},^{2}$\\
$^{1}$Cosmophysics group, Theory Center,
Institute of Particle and Nuclear Studies,
KEK (High Energy Accelerator Research Organization),\\
1-1 Oho, Tsukuba 305-0801, Japan\\ 
$^{2}$Department of Particle and Nuclear Physics, The Graduate
University for Advanced Studies, Tsukuba, 305-0801, Japan}
\begin{document}

\pagerange{\pageref{firstpage}--\pageref{lastpage}} \pubyear{2011}

\maketitle

\label{firstpage}

\begin{abstract}
We investigate the effects of the convection in the hyperaccretion
disk around a stellar-mass black hole, which is considered to be the
central engine of gamma-ray bursts (GRBs), with simple analytical
calculations.  If the convective energy transfer in the vertical
direction becomes efficient compared with the inward advective energy
transport, the hyperaccretion disk is expected to be hotter and the
neutrino emission due to the electron-positron annihilation becomes
the most efficient cooling process.  We find that the sequence of the thermal equilibrium solutions for the convective hyperaccretion disk would have a viscously unstable branch, especially when the viscosity parameter is relatively small ($\alpha \lesssim 0.01$).  This means that the sporadical mass accretion onto a black hole would occur in this disk.  We propose that this process can be the origin of the highly variable lightcurves observed in the prompt emissions of GRBs.
\end{abstract}


\section{Introduction}

Gamma-ray bursts (GRBs) are the most energetic explosions in the
universe, which release large amount of energy up to $10^{51}$
 ergs in only 0.01 -- $10^3$ seconds in the
form of gamma-ray emissions, and they show violent time variabilities
in their emissions ($\delta t\sim 1{\rm msec}$).  Although the central
engine of GRBs is totally hidden from our view, according to their
enormous power, it is often argued that relativistic phenomena should
be involved.  The most popular model for the central engines of GRBs
is a hyperaccreting black hole model (Narayan et al. 1992, 2001).  In
this model, the energy of relativistic jets that makes intense
gamma-ray emissions is produced via the accretion of a massive disk
($0.1-1M_{\odot}$) onto a stellar-mass black hole.  Such systems are
expected after several energetic phenomena such as mergers of double
neutron star binaries (Eichler et al. 1989), neutron star/black hole
binaries (Paczy\'nski 1991), white dwarf/black hole binaries (Fryer et
al. 1999), black hole/He star binaries (Fryer \& Woosley 1998), and
failed supernovae (or collapsars; Woosley 1993; Paczy\'nski 1998;
MacFadyen \& Woosley 1999).  Their mass accretion rate is supposed to
be as high as $0.01-10M_{\odot}{\rm s}^{-1}$, and then the accretion
flows become so optically thick that photons are almost trapped in the
flows. Therefore radiative cooling is not efficient in this case.
Alternatively, since the density and temperature can become very high
($\rho \gtrsim 10^7{\rm g}~{\rm cm}^{-3}$, $T\gtrsim 10^{10}{\rm K}$),
the cooling via thermal neutrino emissions will become efficient.
Such disks (or flows) are called ``Neutrino-Dominated Accretion
Flows'' (NDAFs).  This neutrino emissions may provide the energy
deposition enough to making relativistic jet by neutrino annihilation
above the disk (Popham et al. 1999).  The steady structure of NDAFs
has been studied by many authors (Popham et al. 1999; Narayan et
al. 2001; Kohri \& Mineshige 2002 (hereafter KM02); Di Matteo et
al. 2002; Kohri et al. 2005; Gu et al. 2006: Chen \& Beloborodov 2007;
Liu et al. 2007; Kawanaka \& Mineshige 2007; see also Chapter 10.6 of Kato et al. 2008).  As for their
time-dependent behavior, some authors performed hydrodynamical
simulations including neutrino cooling (Ruffert \& Janka 1999; Janka
et al. 1999; Proga et al. 2003; Lee et al. 2004; Rosswog 2005; Fujimoto et al. 2006;
Setiawan et al. 2006; Nagataki et al. 2007; Shibata et al. 2007; Metzger et al. 2008; Sekiguchi \& Shibata 2010;
Carballido \& Lee 2010; Taylor et al. 2011), and some authors investigated analytically
the disk instabilities which may occur in NDAFs and originate the fast
variability in the prompt emissions of GRBs (Janiuk et al. 2004, 2007;
Masada et al. 2007).

In this study, we investigate the structure and stability of NDAFs by taking into account the convective energy transfer in the vertical direction, which has not been considered in detail in the previous studies.  In the context of standard accretion disks which is radiation pressure-dominated, some authors have inferred the importance of vertical convective motions (Bisnovatyi-Kogan \& Blinnikov 1977; Shakura et al. 1978; Milsom et al. 1994; Blaes \& Socrates 2001; S\c{a}dowski et al. 2011) and some recent studies have shown the importance of convective motions in the hyperaccretion flows (Milosavljevic et al. 2010; Sekiguchi \& Shibata 2010).  Due to the vertical convection, the internal energy stored in the gas would not be advected inward but transported upward, which alters the equation of energy conservation.  We present the equilibrium solutions of convective NDAFs, discuss their stability, and propose the new scenario for the evolution of NDAFs that can account for the violent time variability in the prompt emission of GRBs. 
 
\section{Models}
In the following discussion, we adopt  equations describing the steady structure of a hyperaccretion disk based on Newtonian gravity.  Angular velocity of gas particles can be approximated as Keplerian: $\Omega (r)=(GM/r^3)^{1/2}$, where $M$ is the mass of the black hole.  Expressions for mass conservation, angular momentum conservation, hydrostatic balance and $\alpha$-viscosity are given, respectively, 

\begin{eqnarray}
\dot{M}&=&-2\pi r v_r\Sigma, \label{massconv} \\
-r^3\nu \Sigma \frac{d\Omega}{dr}&=&\frac{\dot{M}}{2\pi}\left[ r^2 \Omega(r)-r_{\rm in}^2 \Omega(r_{\rm in})\right], \label{angmom} \\
\frac{p}{\rho}&=&\Omega(r)^2 H^2, \label{hydrostat} \\
\rho \nu r\frac{d\Omega}{dr}&=&-\alpha p, \label{alphavis}
\end{eqnarray}
where $v_r$, $\Sigma$, $\nu$, $p$ and $H$ denote the radial velocity, surface mass density, kinematic viscosity coefficient, pressure and scale height of the accretion disk, respectively; $\dot{M}$ and $\alpha$ are the mass accretion rate and viscosity parameter.  Here the pressure $p$ is composed of three terms: $p=p_{\rm rad}+p_{\rm gas}+p_{\rm d}$, where $p_{\rm rad}$, $p_{\rm gas}$, $p_{\rm d}$ are the radiation pressure, the gas pressure, the pressure of the degenerate particles, respectively (for their mathematical description see Section 2.1 of KM02).  We set inner radius of the accretion disk, $r_{\rm in}$ to be the radius of the innermost stable circular orbit around a Schwarzschild black hole, $r_{\rm in}=3r_S$, where $r_S=2GM/c^2$ is the Schwarzschild radius.  In addition, we should consider the equation of energy conservation:
\begin{eqnarray}
T\Sigma v_r \frac{ds}{dr}=Q^+-Q_{\nu}^-, \label{energyeq}
\end{eqnarray}
where $Q^+$ and $Q_{\nu}^-$ are the viscous heating rate and neutrino cooling rate per unit surface area.  As for the right hand side,
the energy dissipation rate per unit surface area can be described as
\begin{eqnarray}
Q^+= \frac{3GM\dot{M}}{4\pi r^3}\left(1-\sqrt{\frac{r_{\rm in}}{r}} \right),
\end{eqnarray}
and the neutrino cooling rate $Q^-$ is composed of four terms:
\begin{eqnarray}
Q_{\nu}^-=(\dot{q}_{Ne}+\dot{q}_{e^+ e^-}+\dot{q}_{\rm brems}+\dot{q}_{\rm plasmon})H.
\end{eqnarray}
where $\dot{q}_{Ne}$, $\dot{q}_{e^+ e^-}$, $\dot{q}_{\rm brems}$ and $\dot{q}_{\rm plasmon}$ are the neutrino emissivity due to the electron/positron capture by a nucleon $N$, due to the electron-positron pair annihilation, due to the nucleon-nucleon bremsstrahlung, and the plasmon decay, respectively (see Section 2.3 of Kohri et al. 2005).  The left hand side is often called as the advective cooling rate:
\begin{eqnarray}
Q_{\rm adv}^-\equiv T\Sigma v_r \frac{ds}{dr},
\end{eqnarray}
which means the inward flux of the gas energy along the flow.  Here
$s$ denotes the entropy per unit mass, $s=(s_{\rm rad}+s_{\rm
gas})/\rho$, where $s_{\rm rad}$ and $s_{\rm gas}$ are the entropy
density of the radiation and the gas, respectively (see Section 2.3 of
KM02).  In the following discussions, we approximated $ds/dr$ as
$s/r$, and define the total cooling rate as $Q^- \equiv Q_{\nu}^-
+Q_{\rm adv}^- $. It would be obvious that the radiative cooling is
negligible.

In order to see effects of convection in an hyperaccretion disk, we should estimate a typical timescale for the convective motion of the blob in the vertical direction $t_{\rm conv}$ and compare it with  timescale for the inward advection $t_{\rm adv}$.  The acceleration of the blob in the accretion disk due to the vertical buoyancy would be the same order of magnitude as that due to the gravitational force, i.e. $\sim \Omega (r)^2 z$ where $z$ is the vertical coordinate measured from the equatorial plane.  Then the convective speed along the vertical direction comming from $z_{\rm min}~(\ll z)$ should be
\begin{eqnarray}
v(z)&\simeq &\sqrt{2\int_{z_{\rm min}}^z dz g(z)} \nonumber \\
&\simeq &\Omega (r) z,
\end{eqnarray}
and then the timescale for the blob which is emerged at $z(<H)$ and transported convectively can be estimated as
\begin{eqnarray}
t_{\rm conv}&\simeq &\int_z^H \frac{dz^{\prime}}{v(z^{\prime})} \nonumber \\
&\simeq & \frac{1}{\Omega (r)}\ln \left( \frac{H}{z} \right).
\end{eqnarray}

On the other hand, advection timescale can be estimated as
\begin{eqnarray}
t_{\rm adv}&\simeq &\frac{r}{v_r} \nonumber \\
&\simeq &\frac{1}{\alpha \Omega(r)} \left( \frac{r}{H} \right) ^2.
\end{eqnarray}
Here we concentrate on cases with $\dot{M}\lesssim 10^{-3}M_{\odot}{\rm sec}^{-1}$.  According to KM02 etc., such an accretion disk is very optically thick with respect of photons and not so dense or hot enough to emit neutrinos efficiently.  However, when the convection timescale is shorter than the advection timescale, the energy dissipated in the accretion flow would not be advected radially inward, but would be convected upward, and as a result the gas
 at a certain radius would be heated more efficiently than in the case without convection.  Let us evaluate structures of such accretion flows with simple analytical discussions.  The convective motion can prevent the advective energy transport when $t_{\rm conv}<t_{\rm adv}$, which would be fulfilled when $z>z_c$ where $z_c$ should satisfy
\begin{eqnarray}
\frac{z_c}{H}=\exp\left( -\frac{1}{\alpha}\left( \frac{r}{H} \right)^2
\right).
\label{eq:zmH}
\end{eqnarray}
Therefore, the equation of energy conservation should be modified by
multiplying the factor $z_{c}/H$ to the left-hand side of Eq.~(\ref{energyeq}) to be
\begin{eqnarray}
T\Sigma \frac{z_c}{H} v_r \frac{ds}{dr} =Q^+ - Q_{\nu}^-.
\end{eqnarray}
Other effects should be much milder than this suppression as will be
discussed later.

Because of this suppression of the advective cooling, the significant amount of internal energy of the gas stay at the same radial position in the accretion disk.  As a result the disk temperature would be much higher than in the case when the convection is neglected, and then the disk structure is expected to be significantly changed.  In order to see this, we plot the sequence of the thermal equilibrium solutions for convective NDAFs on the $(\Sigma, \dot{M})$ and $(\Sigma, T)$ plane for various values of $\Sigma$, fixing the radial position $r$.
 
\section{Stability of convective NDAFs}
By finding the sets of physical parameters which satisfy the fundamental equations for the disk structure, (\ref{massconv}), (\ref{angmom}), (\ref{hydrostat}) and (\ref{energyeq}), we can derive the sequence of the thermal equilibrium solutions.  In Fig. 1 we show them on the $(\Sigma, \dot{M})$ plane for $r=4r_S$, varying the viscosity parameter $\alpha$.  Here the mass accretion rate is normalized by the critical mass accretion rate,
\begin{eqnarray}
\dot{M}_{\rm crit}=16L_{\rm Edd}/c^2,
\end{eqnarray}
where $L_{\rm Edd}$ is the Eddington luminosity,
\begin{eqnarray}
L_{\rm Edd}=4\pi GM_{\rm BH}m_p c/\sigma_T,
\end{eqnarray}
with proton mass $m_p$ and the Thomson cross section $\sigma_T \simeq 6.6\times 10^{-25}{\rm cm}^2$.  We can see that for lower values of $\alpha$ the thermal equilibrium curves have a part with negative gradient, i.e.
\begin{eqnarray}
\left( \frac{d\dot{M}}{d\Sigma} \right)_{Q^+=Q^-}<0,
\end{eqnarray}
which means that this accretion flow is dynamically unstable (see,
e.g. Lightman \& Eardley 1974; Kato et al. 2008).  In such accretion
disks, we can expect the sporadical mass accretion onto the central
black hole, which may occur in dwarf novae and some microquasars with
a radiation pressure-dominated accretion disk.  This can be the source
of violent time variability in the GRB prompt emissions with the
mass accretion rate of a few times $10^{-4}M_{\odot}{\rm s}^{-1}$, which corresponds to $10^{50}-10^{51}$ erg~s$^{-1}$.

\begin{figure}
\begin{center}
\includegraphics[width=80mm]{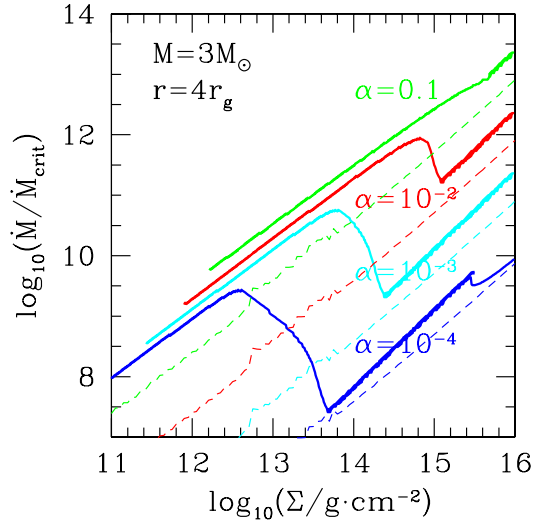}
\end{center}
\caption{Thermal equilibrium curves of convective hyperaccretion flows
on the $(\Sigma,\dot{M})$ plane at $r=4r_S$ (solid lines) with
$\alpha=10^{-1}$ ({\it green line}), $10^{-2}$ ({\it red line}),
$10^{-3}$ ({\it cyan line}) and $10^{-4}$ ({\it blue line}), plotted
from top to bottom.  The central black hole mass is set to be
$3M_{\odot}$.  The thermal equilibrium curves without taking into
account the convection are also shown (dashed lines). Right regions
with respect to the bending part are dominated by degenerate-electron
pressure in which we are not interested in the current study.}

\label{f1}

\end{figure}

We also show in Fig. 2 the thermal equilibrium curves on the $(\Sigma, T)$ plane with the same parameter sets as in Fig. 1.  Here we can see that in the whole parameter space the thermal stability criterion,
\begin{eqnarray}
\left( \frac{dQ^+}{dT} \right)_{\Sigma} < \left( \frac{dQ^-}{dT} \right)_{\Sigma},
\end{eqnarray}
is satisfied, and so this accretion flow is thermally stable. 

\begin{figure}
\begin{center}
\includegraphics[width=80mm]{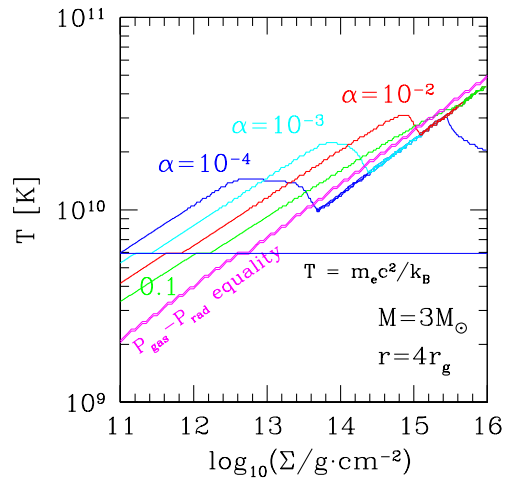}
\end{center}
\caption{Thermal equilibrium curves on the $(\Sigma, T)$ plane at $r=4r_S$ with the same parameter sets as in Fig. 1.  There is a notable
difference between the current case and the convection-inefficient one
where the $T-\Sigma$ relation did not depend on $\alpha$ in the normal
slim-disk or NDAF solutions (see Table 1 of KM02).}

\label{f2}

\end{figure}

\section{Discussion}
Here we discuss the physical origin of the dynamical instability of NDAFs with vertical convective motions presented in the last section.  To see this let us express the physical properties of the accretion disk in terms of ($\Sigma$, $\dot{M}$) or ($\Sigma$, $T$).  As for the former set of variables, it is obvious from Eq. (6) that the viscous heating rate $Q^+$ is proportional to $\dot{M}$.  We can also express $Q^+$ with the latter set of variables as follows.  In the accretion flows with a mass accretion rate much larger than $\dot{M}_{\rm crit}$, the pressure and entropy are dominated by radiation, i.e. $p_{\rm rad} \gg p_{\rm gas}, p_{\rm d}$ and $s_{\rm rad} \gg s_{\rm gas}$.  In this case, the scale height $H$ would be proportional to $T^4 \Sigma ^{-1} \Omega^{-2}$.  Therefore, the viscous heating rate $Q^+\approx 2\alpha p H \Omega$ should be proportional to $\alpha T^8 \Sigma ^{-1} \Omega^{-1}$.

On the other hand, as for the neutrino cooling processes, the emissivity by the electron/positron capture process in electron-nondegeneracy limit can be approximated as
\begin{eqnarray}
\dot{q}_{Ne}=9.2\times 10^{33}~{\rm ergs}~{\rm cm}^{-3}~{\rm s}^{-1} \left(\frac{T}{10^{11}{\rm K}} \right)^6 \left(\frac{\rho}{10^{10} {\rm g}~{\rm cm}^{-3}} \right),
\end{eqnarray}
while the emissivity by the electron-positron annihilation is
\begin{eqnarray}
\dot{q}_{e^{+}e^{-}}=4.8\times 10^{33}~{\rm ergs}~{\rm cm}^{-3}~{\rm s}^{-1} \left( \frac{T}{10^{11}~{\rm K}} \right) ^9,
\end{eqnarray}
in the nondegeneracy limit.  In the unstable branch found in Fig. 1 (and the corresponding part in Fig. 2), we can see that the pair annihilation process dominates the electron/positron capture process.  Therefore $Q_{\nu}$ is approximately proportional to $T^9 H \propto T^{13}\Sigma^{-1}$.

By equating the viscous heating rate and the neutrino cooling rate,
 $Q^+$ and $Q_{\nu}^-$, we can easily see that the temperature $T$ is
  proportional to $\alpha^{1/5} M^{1/15} r^{-3/10}$ and does not depend on $\Sigma$, which is also found in Fig. 2, because both $Q^+$ and $Q_{\nu}^-$ are proportional to $\Sigma^{-1}$ when described in terms of ($\Sigma$, $T$).  Considering the same part of the thermal equilibrium curves on the ($\Sigma$, $\dot{M}$) plane, as $\dot{M} \propto \Omega^{-2} Q^+ \propto T^8 \Sigma^{-1}$, we can see that $\dot{M} \propto \alpha^{13/8} M^{-7/10} r^{21/10} \Sigma^{-1}$, which is seen in Fig. 1 as a negative gradient part in each thermal equilibrium curve.

We can interpret the physical reason of the viscously unstable branch in the convective NDAFs as follows; due to the vertical convection in the accretion disk, the advective energy transport would be significantly suppressed, and thus the disk temperature would become higher than in the case without the convection.  In such a hot disk, electrons would not be degenerate and hence the electron-positron annihilation process would become an efficient cooling process.  It can dominate the emissivity due to the electron/positron capture by nucleons ($\propto T^6 \rho$), which was the dominant neutrino emission process in the NDAF models without the convection (e.g. see KM02).  Therefore the temperature-dependence of the cooling rate would be changed and make the unstable branch appear in the ($\Sigma$, $\dot{M}$) plane.  We can see in Fig. 1 that with smaller viscous parameter $\alpha$ the unstable branch appears with smaller mass accretion rate $\dot{M}$.  This is because the advective energy transport is more inefficient with smaller $\alpha$, which makes the disk temperature high enough for the electron-positron annihilation process to become dominant even with smaller mass accretion rate.

In particular, the typical timescale for this instability is the order of the viscous timescale,
\begin{eqnarray}
t_{\rm vis}&\sim &\frac{1}{\alpha \Omega} \left(\frac{r}{H}\right) ^2 \nonumber \\
&\sim &4\times 10^{-2}~{\rm sec} \left( \frac{\alpha}{0.01} \right)^{-1} \left( \frac{M}{3M_{\odot}} \right)^{-1/2} \nonumber \\
&&\times \left( \frac{r}{4\times 10^6{\rm cm}} \right)^{3/2} \left(\frac{r}{H}\right) ^2,
\end{eqnarray}
and therefore when $r/H \lesssim {\cal O}(1)$, which is often realized in the innermost region of NDAF, this timescale is short enough to account for the time variability observed in the prompt emissions of GRBs.

\section{Conclusion and Summary}
In this study, we show that if the convective energy transfer along the vertical direction, which is naturally expected in the radiation-pressure dominated accretion disk, the thermal equilibrium solution for a hyperaccretion disk would have a viscously unstable branch.  When this branch is realized, highly time-dependent mass accretion onto a black hole would occur.  Such instabilities have been studied in the contexts of dwarf novae (Meyer \& Meyer-Hofmeister 1981; Hoshi 1979; Smak 1982) and microquasars which have radiation pressure-dominated accretion disks in themselves (Lightman \& Eardley 1974; Shibazaki \& Hoshi 1975; Pringle 1976; Shakura \& Sunyaev 1976; Abramowicz et al. 1988), and the time variable features observed in those sources are theoretically interpreted with the model of those instabilities.  In the context of GRBs, the process discussed in this study may lead to the intermittent energy release from a hyperaccretion disk, and then the highly inhomogeneous jet would be launched, which can be the origin of the short-term variability in the prompt emission of GRBs.  Judging from isotropic luminosities of observed long GRBs ($L_{\rm iso} \sim 10^{51-52}
{\rm erg}~{\rm s}^{-1}$), the accretion rate adopted here $\dot{M}\sim 10^{-3}-10^{-4}M_{\odot}{\rm s}^{-1}\sim 10^{50-51}{\rm erg}~{\rm s}^{-1}/c^2$ is relatively small.  However, recent observations show that there is a population of low luminosity GRBs ($L_{\rm iso}\sim 10^{46-49}{\rm erg}~{\rm s}^{-1}$; e.g. Galama et al. 1998; Campana et al. 2006), and the variabilities of their prompt emissions may be explained with our model.  Our model may be also appropriate to describing the late time activity of GRBs such as
X-ray flares in afterglows (e.g. Burrows et al. 2005).

We should note that the absorption optical depth for neutrinos is smaller than unity, according to  Eqs. (55)-(57) in Kohri et al. (2005).  This means that we do not have to consider the neutrino trapping effect in the current situation ($T\lesssim 10^{11}{\rm K}$, $\Sigma \lesssim 10^{15}{\rm g}~{\rm cm}^{-2}$).

We should mention the effect of magnetohydrodyanmic turbulence on the convection in our accretion disks.  A hyperaccretion
disk should have a turbulent structure due to the magnetorotational
instability (MRI), and it would help the mass accretion as the turbulent
viscosity (cf. Balbus and Hawley 1991).  Here, using $\alpha$ parameter, we can describe the turbulent viscosity $\nu_t$ as $\nu_t\sim \alpha c_s H$ where $c_s$ and $H$ are the speed of sound and the scale height of an accretion flow, respectively.  In our study $\alpha$ parameter is assumed to be relatively small value, $\lesssim 10^{-2}$, which may be due to the suppression of MRI in a hyperaccretion flow (Masada et al. 2007).  In other words, the typical scale and the speed of MRI turbulence in our accretion flows are much smaller than the scale height and the speed of sound, respectively.  So, in our cases, we can say that the convective motion would not be disturbed by MRI turbulence because the typical scale length and speed of convection are $\sim H$ and $\sim c_s$, respectively.

The convective energy transport along the vertical direction can also
affect the structure of the slim disk, in which the optical depth for
photon $\tau$ is so large that the photon trapping is efficient.  If
the timescale for the vertical convection is shorter than that of
radiative diffusion, the photons generated in the equatorial plane can
be convected upward to the disk surface and emitted out efficiently.
This means that the effective optical depth would get much smaller
than that in the case without convection. This modification should appear
in the radiative cooling term of the energy conservation equation approximately by multiplying a factor of
$z_{c}/H$, to the optical depth, i.e.,  like $\tau_{\rm eff} \sim \tau
\times z_{c}/H$.  This modification in $\tau$ makes the equilibrium
curve for the standard disk shift to a larger $\Sigma$ and a higher
$T$ regions such as $\Sigma \sim 10^{4} - 10^{6}{\rm g}~{\rm cm^{-2}}$
and $T \sim 10^{8} K$, respectively. In addition to these effects in
the multi-dimensional treatments, we should also consider a diffusion
of photon along with the $z$-axis, which also affects the
photon-trapping region (Kohri et al. 2007). Therefore although the
current NDAF region with $\Sigma \sim 10^{13}-10^{15}{\rm g}~{\rm
cm^{-2}}$ and $T \sim 10^{10}K$ discussed in this letter should not be
affected by these modifications, to know the whole structure of the
disks, we would have to perform much more precise numerical
simulations with reliable treatments of the convection, the diffusion
and the neutrino transfer.

\ \\

We thank Shin Mineshige and Hiroki Nagakura for useful comments.  This work is supported by the World Premier International Center Initiative
(WPI Program), MEXT, Japan and the Grant-in-
Aid for Science Research, Japan Society for the Promotion
of Science (No. 22740131 for NK; No. 22244030, 21111006 for KK), and K.K. is supported by the Center for the Promotion of
Integrated Sciences (CPIS) of Sokendai.

\label{lastpage}
\end{document}